# Charge writing at the LaAlO$_3$/SrTiO$_3$ surface

Yanwu Xie,[†,‡] Christopher Bell,[†,§] Takeaki Yajima,[†] Yasuyuki Hikita,[†] and Harold Y. Hwang[†,§,]*

[†]Department of Advanced Materials Science, University of Tokyo, Kashiwa, Chiba 277-8561, Japan

[‡]State Key Laboratory of Metastable Materials Science and Technology, Yanshan University, 066004, Qinhuangdao, P. R. China

[§]Japan Science and Technology Agency, Kawaguchi, 332-0012, Japan

*To whom correspondence should be addressed, email: hyhwang@k.u-tokyo.ac.jp

ABSTRACT

Biased conducting-tip atomic force microscopy (AFM) has been shown to write and erase nanoscale metallic lines at the LaAlO$_3$/SrTiO$_3$ interface. Using various AFM modes, we show the mechanism of conductivity switching is the writing of surface charge. These charges are stably deposited on a wide range of LaAlO$_3$ thicknesses, including bulk crystals. A strong asymmetry with writing polarity was found for 1 and 2 unit cells of LaAlO$_3$, providing experimental evidence for a theoretically predicted built-in potential.

Atomic force microscopy (AFM) is an essential tool for characterizing solid surfaces[1-3] and modifying materials at the nanometer scale. Examples include poling ferroelectric thin films,[4] electro-chemical modification,[5] surface atom manipulation,[6] electret charging for data storage[7] or nanoxerography,[8] and probing two-dimensional electron systems.[9,10] Recently, biased conducting-tip AFM was shown to write and erase nanoscale metallic lines at the LaAlO$_3$/SrTiO$_3$ interface,[11,12] a system notable for interface conductivity between two insulators.[13] This is intriguing in that surface modification remotely toggles a proximate interface. Here we show the mechanism of switching interface conductivity is the writing of surface charge, as established by various AFM modes. These charges are stably deposited on a wide range of LaAlO$_3$ thicknesses, including bulk crystals. A strong asymmetry with writing polarity was found for 1 and 2 unit cells of LaAlO$_3$, providing experimental evidence for a theoretically predicted built-in potential.[14-16] These results suggest that thin LaAlO$_3$ films can be utilized for nanoscale charge manipulation in a wide range of oxide heterostructures.

A key result in studying the LaAlO$_3$/SrTiO$_3$ (LAO/STO) interface demonstrated a critical LAO thickness, $d_{LAO}$ = 4 unit cells (uc), below which the interface abruptly changes from metallic to insulating.[17] This was interpreted as an electronic reconstruction to resolve the polar discontinuity between STO and LAO.[14-16,18,19] It was also shown that by applying a gate voltage across the STO substrate, a $d_{LAO}$ ~ 3 uc film could be reversibly driven conducting.[17] To achieve this on the nanoscale, a biased AFM tip was applied to the LAO surface, inducing conductivity.[11,12] In these experiments, conductivity changes were also the diagnostic probe, since no topographic differences were found except at the highest voltages.[11] While large conductivity changes were found for $d_{LAO} \sim 3.3$ uc, none occurred for thinner films,[11] and for $d_{LAO} \geq 4$ uc the interface is already highly metallic,[17] masking any effects of AFM writing. Here we use electric force microscopy (EFM), surface potential and apparent height measurements to determine the charge nature of the surface modification responsible for the nanoscale interface wires.

The LAO/STO samples were prepared by depositing LAO layers with thicknesses ranging from 1 to 25 uc on $TiO_2$-terminated (001) STO substrates. All AFM experiments were carried out using a multimode AFM with a conducting tip, in air, at room temperature. For writing, we apply a bias to the tip and utilize a relatively low amplitude set point (1 to 3 % of the typical imaging value) to approach close to the surface. For reading in EFM mode, the AFM was operated by oscillating the cantilever at its resonant frequency (~270 kHz) while scanning at a lift height of 20 nm. A schematic of the measurement configuration is shown in Figure 1f. The smooth step and terrace surface of the LAO/STO samples enables the electric field variation to be clearly detected, avoiding cross-talk from the surface topography. Before the conductivity switching experiment, the sample was kept in a dark environment for more than 10 hours to suppress photo-excited carriers in the bulk $SrTiO_3$. Further details are given in the Supporting Information.

First, we reproduced the reversible conductivity changes previously reported,[11,12] and show that they can be detected by EFM. Initially, two electrodes were written ($V_{write}$ = +8 V) on a $d_{LAO}$ = 3 uc sample, connecting to Al electrodes and an external circuit (Figure 1a). Subsequently, the electrodes were connected ($V_{write}$ = +6 V, Figure 1b), resulting in a sharp conductance increase (label 1, Figure 1e). Next, the conductive path was cut (Figure 1c) by writing perpendicular to the wire ($V_{write}$ = -3 V), giving a sharp conductance decrease (label 2, Figure 1e). This process is reproducible: reconnection was made writing vertically ($V_{write}$ = +6 V, label 3, Figure 1e). Using EFM phase images, we map the surface fields corresponding to the conductive paths (dark contrast, reading tip voltage $V_{read}$ = -2 V) and the insulating paths (bright contrast), which directly correlate with the resistance changes. These results indicate that the conductivity changes induced by the AFM for $d_{LAO}$ = 3 uc are due to writing surface charge.

To determine the sign of this surface charge, Figures 2a and b show EFM phase images for two diamond patterns written with $V_{write}$ = +5 V (inside) and -5 V (outside), on a $d_{LAO}$ = 10 uc sample. These data were captured with $V_{read}$ = +2 V and -2 V, respectively, and show inverted contrast. In EFM, an attractive force gradient reduces the cantilever resonant frequency, resulting in a negative shift (darker) in the phase image, while a repulsive force gradient gives a positive shift (brighter).[4,20] The phase change direction thus depends on the signs of both $V_{write}$ and $V_{read}$: *e.g.* positive charges give a positive phase shift for $V_{read}$ > 0. This relationship (as well as the scaling with LAO thickness discussed below) also indicates that the field measured arises from the surface charge, and not from the electrons at the buried interface. We further measured the surface potential (lift height = 20 nm), of the same patterns (Figure 2c), showing that $V_{write}$ > 0 leads to a positive potential, while $V_{write}$ < 0 gives a negative potential. Finally Figure 2d shows the apparent height measured in tapping mode with $V_{read}$ = -5 V. In this case an attractive electrostatic force gives an increase in the effective height, and the reverse for repulsion.[20,21] Thus all of these methods (and for all samples studied) confirm that writing with $V_{write} > 0$ accumulates positive surface charge, and the reverse for $V_{write} < 0$.

To examine the bias dependence of this charge writing, Figure 3 shows EFM phase images from a 5 μm × 5 μm region of a $d_{LAO}$ = 10 uc sample, where nine lines have been written ($V_{write}$ varied from -8 V to +8 V in 2 V steps). The resulting EFM phase images, taken with $V_{read}$ = +2 V and -2 V, are shown in Figure 3a and 3b, respectively. There were no measurable phase shifts outside of the written areas, nor where $V_{write}$ = 0 V. Conventional topographic images (Figure 3c) showed no change in the written area, except for the $V_{write}$ = -8 V line, where slight surface damage is visible. We observe a strong asymmetry in the response: $V_{write} > 0$ gives a smaller phase change than $V_{write} < 0$ (Figure 3d). Further insight can be gained by examining the history and bias dependence of the features using the same $d_{LAO}$ = 10 uc sample. The same line was repeatedly drawn with $V_{write}$ following the sequence (in units of V): +4,-4,+6,-6,+8,-8,+8. Large hysteresis was observed, as well as the asymmetry with tip bias (Figure 4a).

Figure 4b compares the observed phase shifts in a series of samples with 1 uc ≤ $d_{LAO}$ ≤ 25 uc. All data were collected using only two AFM tips (one for negative writing, another for positive writing). Despite some scatter associated with variation and degradation of the tips, we find that as $d_{LAO}$ decreases, the phase shift induced by $V_{write} < 0$ decreases, sharply at 2 uc. By contrast, the phase shift for features written by $V_{write} > 0$ varies only slightly for $d_{LAO}$ ≥ 2 uc, showing an unusually strong response for $d_{LAO}$ = 1 uc. For $d_{LAO}$

= 1 and 2 uc, features written with $V_{write} > 0$, show unstable phase signals, decaying significantly within minutes. By contrast, the written charges for $d_{LAO} \geq 3$ uc are robust for many hours, although a clear decay is observed.

Figure 4c shows the time dependence of the peak phase change of features written on a $d_{LAO} = 10$ uc sample, for $V_{write} < 0$ and $V_{write} > 0$. The data are well fit by exponential decay (solid lines) of the form $p(t) = p_0 + p_1 exp(-t/\tau)$. In both cases $\tau \sim 2000$ s (full fitting parameters are in the Supporting Information), suggesting the same decay mechanism independent of the sign of $V_{write}$. $p_0$ for $V_{write} < 0$ ($V_{write} > 0$) is ~ 60% (30%) of $p(t=0)$, demonstrating the feature stability (> 20 hours in air). The time dependence of the phase profiles (corresponding to Figure 4c), are shown in Figure 4d. For either $V_{write}$ polarity, the width of the profile increases as the peak height decreases, while the total area remains constant within experimental error, indicating lateral diffusion, not compensation, of the charges.[22]

In considering the origin of this surface charge, the hysteresis shown in Figure 4a suggests ferroelectricity, although similar hysteresis has also been observed in metal-nitride-oxide-silicon (MNOS) charge storage devices.[23] The possibility of STO ferroelectricity[24] can be excluded since this predicts the opposite sign of conductance change, and a weaker response for thicker LAO would be expected, contrary to the observations (Figure 4b). So although the STO at the LAO/STO interface is known to be strongly polarized,[25] we conclude it is not switchable. Ferroelectricity in LAO could be consistent with our data, similar to observation of overscreening the ferroelectric polarization in PZT films.[4] Indeed, several theoretical studies indicate that LAO polarization is a key determinant of the critical thickness.[14-16] However, we have found robust charge writing on bare LAO substrates (see Supporting Information). This was much more stable than writing on STO which decays relatively rapidly.[26] Given that bulk LAO is clearly not ferroelectric, we also exclude this mechanism.

We are left with the conclusion that written charges are on or very near the surface. This is supported by the observation that these features can be easily removed by gently wiping the LAO surface using acetone. While the microscopic form of this surface charge remains an open question, we comment on three scenarios. Firstly, it is well known that the biased tip may desorb surface adatoms and/or dissociate surface atoms. Indeed, the removal of oxygen from the LAO surface was suggested for $V_{write} > 0$ and recovery for $V_{write} < 0$.[11] However, the phase contrast for $V_{write} < 0$ observed here implies adding excess oxygen into LAO, which is unlikely. Secondly, direct injection of electrons onto the LAO surface could explain all features of our data, as is implicitly believed to occur for AFM-tip-induced charges in many semiconductor structures. However, the robustness of the surface charge is surprising in comparison to GaAs/AlGaAs two-dimensional electron gas modulation, for example, which is only stable at low temperatures in cryogenic vacuum.[9] Thirdly, we note that AFM-tip-induced water-ion injection was proposed to explain surface charges formed on insulating polymers.[27,28] The robustness of the features observed here suggest that field-ionization of water may be important to create the charges we have observed and manipulated.

Independent of the form of the surface charge, a number of our observations fall within the polar discontinuity framework, where a negative potential (relative to the grounded interface) on the surface of unreconstructed LAO is predicted.[14-16,18,19] This potential would facilitate the accumulation of positive charges and hinder negative charges, which is mirrored in the strongly asymmetric response of the 1 and 2 uc layers (Figure 4b): writing with $V_{write} > 0$ shows a greatly enhanced response compared to $V_{write} < 0$. This surface potential tends to adsorb atmospheric molecules, *e.g.* water, which facilitate the diffusion of the charges, consistent with the fast decay for $d_{LAO}$= 1 and 2 uc. These features suggest that the surface writing technique discovered by Cen *et al.*[11] has the potential to be extended to many other heterostructures in close proximity to the LAO surface,[29,30] allowing for the broad ability to manipulate electronic structure on the nanoscale.

**References**

(1) Binnig, G.; Quate, C. F.; Gerber, Ch. *Phys. Rev. Lett.* **1986**, 56, 930-933.

(2) Giessibl, F. J. *Science* **1995**, 267, 68-71.

(3) García, R.; Pérez, R. *Surf. Sci. Rep.* **2002**, 47, 197-301.

(4) Ahn, C. H.; Tybell, T.; Antognazza, L.; Char, K.; Hammond, R. H.; Beasley, M. R.; Fischer, Ø.; Triscone, J. M. *Science* **1997**, 276, 1100-1103.

(5) Wouters, D.; Schubert, U. S. *Angew. Chem. Int. Ed.* **2004**, 43, 2480-2495.

(6) Custance, O.; Perez, R.; Morita, S. *Nature Nanotech.* **2009**, 4, 803-810.

(7) Barrett, R. C.; Quate, C. F. *Ultramicroscopy* **1992**, 42-44, 262-267.

(8) Mesquida, P.; Stemmer, A. *Adv. Mater.* **2001**, 13, 1395-1398.

(9) Crook, R.; Graham, A. C.; Smith, C. G.; Farrer, I.; Beere, H. E.; Ritchie, D. A. *Nature* **2003**, 424, 751-754.

(10) Tessmer, S. H.; Glicofridis, P. I.; Ashoori, R. C.; Levitov, L. S.; Melloch, M. R. *Nature* **1998**, 392, 51-54.

(11) Cen, C.; Thiel, S.; Hammerl, G.; Schneider, C. W.; Andersen, K. E.; Hellberg, C. S.; Mannhart, J.; Levy, J. *Nature Mater.* **2008**, 7, 298-301.

(12) Cen, C.; Thiel, S.; Mannhart, J.; Levy, J. *Science* **2009**, 323, 1026-1030.

(13) Ohtomo, A.; Hwang, H. Y. *Nature* **2004**, 427, 423-426.

(14) Popovic, Z. S.; Satpathy, S.; Martin, R. M. *Phys. Rev. Lett.* **2008**, 101, 256801.

(15) Lee, J.; Demkov, A. A. *Phys. Rev. B* **2008**, 78, 193104.

(16) Pentcheva, R.; Pickett, W. E. *Phys. Rev. Lett.* **2009**, 102, 107602.

(17) Thiel, S.; Hammerl, G.; Schmehl, A.; Schneider, C. W.; Mannhart, J. *Science* **2006**, 313, 1942-1945.

(18) Nakagawa, N.; Hwang, H. Y.; Muller, D. A. *Nature Mater.* **2006**, 5, 204-209.

(19) Hwang, H. Y. *Science* **2006**, 313, 1895-1896.

(20) Guillemot, C.; Budau, P.; Chevriver, J.; Marchi, F.; Comn, F.; Alandi, C.; Bertin, F.; Buffet, N.; Wyon, Ch.; Mur, P. *Europhys. Lett*. **2002**, 59, 566-571.

(21) Yan, M.; Bernstein, G. H. *Ultramicroscopy* **2006**, 106, 582-586.

(22) Buh, G. H.; Chung, H. J.; Kuk, Y. *Appl. Phys. Lett*. **2001**, 79, 2010-2012.

(23) Frohman-Bentchkowsky, D.; Lenzlinger, M. *J. Appl. Phys.* **1969**, 40, 3307-3319.

(24) Haeni, J. H.; Irvin, P.; Chang, W.; Uecker, R.; Reiche, P.; Li, Y. L.; Choudhury, S.; Tian, W.; Hawley, M. E.; Craigo, B.; Tagantsev, A. K.; Pan, X. Q.; Streiffer, S. K.; Chen, L. Q.; Kirchoefer, S. W.; Levy, J.; Schlom, D. G. *Nature* **2004**, 430, 758-761.

(25) Ogawa, N.; Miyano, K.; Hosoda, M.; Higuchi, T.; Bell, C.; Hikita, Y.; Hwang, H. Y. *Phys. Rev. B* **2009**, 80, 081106.

(26) Uchihashi, T.; Okusako, T.; Yamada, J.; Fukano, Y.; Sugawara, Y.; Igarashi, M.; Kaneko, R.; Morita, S. *Jpn. J. Appl. Phys.* **1994**, 33, L374-376.

(27) Knorr, N.; Rosselli, S.; Miteva, T.; Nelles, G. *J. Appl. Phys.* **2009**, 105, 114111.

(28) Rezende, C. A.; Gouveia, R. F.; da Silva, M. A.; Galembeck F. *J. Phys.: Condens. Matter* **2009**, 21, 263002.

(29) Higuchi, T.; Hotta, Y.; Susaki, T.; Fujimori, A.; Hwang, H. Y. *Phys. Rev. B* **2009**, 79, 075415.

(30) Takizawa, M.; Hotta, Y.; Susaki, T.; Ishida, Y.; Wadati, H.; Takata, Y.; Horiba, K.; Matsunami, M.; Shin, S.; Yabashi, M.; Tamasaku, K.; Nishino, Y.; Ishikawa, T.; Fujimori, A.; Hwang, H. Y. *Phys. Rev. Lett.* **2009**, 102, 236401.

**Acknowledgements.** We thank C. Stephanos for useful discussions, and M. Lippmaa for the use of clean room facilities. Y.W.X. acknowledges funding from the Japan Society for the Promotion of Science (JSPS).

**Supporting Information Available**. Materials and methods, writing and erasing conductive paths on the $d_{LAO}$ = 3 uc sample, AFM parameters, exponential decay fitting, and charge writing on bare (001) LAO and (001) STO substrates. This material is available free of charge via the Internet at http://pubs.acs.org.

**Figure legends:**

FIGURE 1. Writing and erasing conductive paths for $d_{LAO}$ ~ 3 uc. EFM phase images with $V_{read}$ = -2 V. (a) Dark contrast shows the electrodes connected to the outer Al wiring, $V_{write}$ = +8 V. (b) The written wire connects the electrodes, $V_{write}$ = +6 V, corresponding to (1) in (e). (c) The wire is cut (bright phase contrast) with $V_{write}$ = -3 V, corresponding to (2) in (e). (d) Electrodes reconnected with a vertical line using $V_{write}$ = +6 V, corresponding to (3) in (e). (e) Two-point conductance between the Al electrodes. (f), Schematic for AFM writing.

FIGURE 2. Determining the sign of the written charge. (a, b) EFM phase images of two diamond patterns written on a $d_{LAO}$ = 10 uc sample, $V_{write}$ = +5 V (inside) and -5 V (outside), acquired with $V_{read}$ set to (a), +2 V and (b), -2 V. (c) Surface potential images of the same patterns, at a lift height of 20 nm (the weak background streaks seen here are artifacts from temporally coherent noise, and change with the tip scan rate). (d) Apparent topography with $V_{read}$ = -5 V for the same patterns.

FIGURE 3. Voltage dependence of charge writing on a $d_{LAO}$ = 10 uc sample. (a, b) EFM phase image of a 5 µm × 5 µm area after writing nine lines varying $V_{write}$ from -8 V (top) to +8 V (bottom), read with $V_{read}$ = +2 V (a), and $V_{read}$ = -2 V (b). (c) Topographic image of the same area as in (a) and (b). The surface is smooth with a clear 1 uc step-and-terrace structure. (d) Line profiles across images (a) and (b). The signals were averaged over the written regions.

FIGURE 4. Bias, thickness and time dependence of the EFM phase change. (a) Phase hysteresis loops versus $V_{write}$ on a $d_{LAO}$ = 10 uc sample ($V_{read}$ = +5 V). (b) Variation of the phase switch with $d_{LAO}$. Three data sets acquired with the same tip and tip settings are shown for both $V_{write}$ = -6 V (closed symbols) and $V_{write}$ = +8 V (open symbols), $V_{read}$ = -2 V. All data are measured within ~ 90 s of the end of writing. For $V_{write}$ = +8 V, the phase changes in the $d_{LAO}$ = 1 and 2 uc samples (green) decay significantly within minutes, in contrast to the thicker films (blue) which are relatively stable. Lines are guides to the eye. (c) Time-dependence of the peak phase switch of the written features on a $d_{LAO}$ = 10 uc sample. Solid lines are fits to exponential decay. (d) Profiles across the written lines after various intervals.

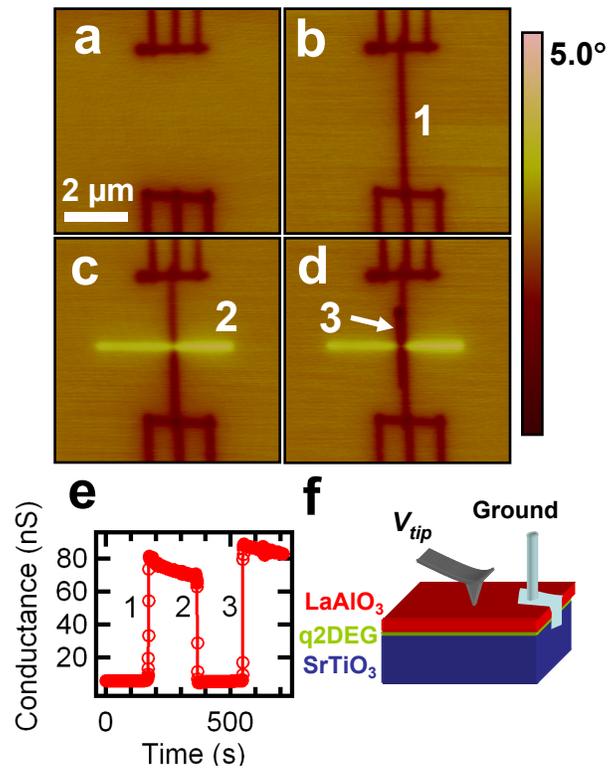

Figure 1  Y. W. Xie *et al.*

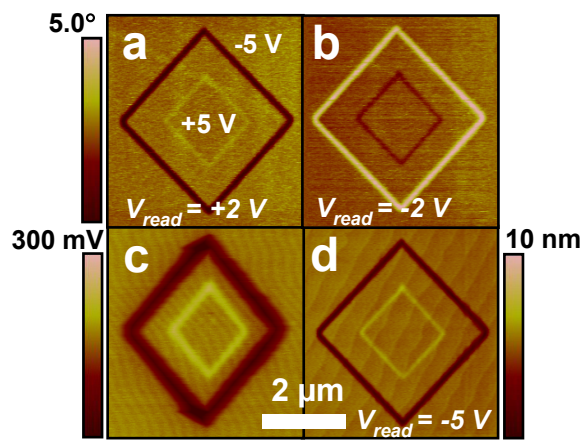

Figure 2  Y. W. Xie *et al.*

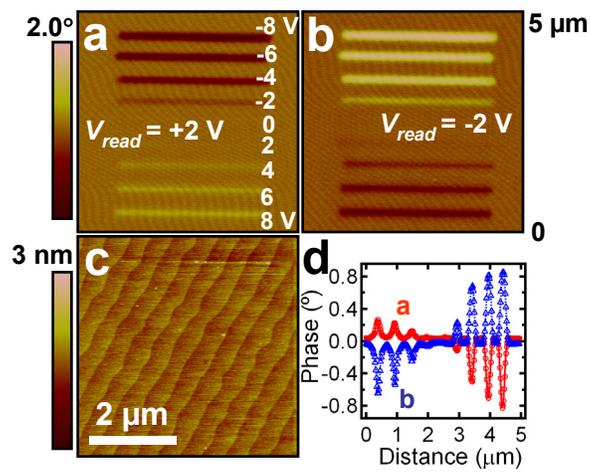

Figure 3 Y. W. Xie *et al.*

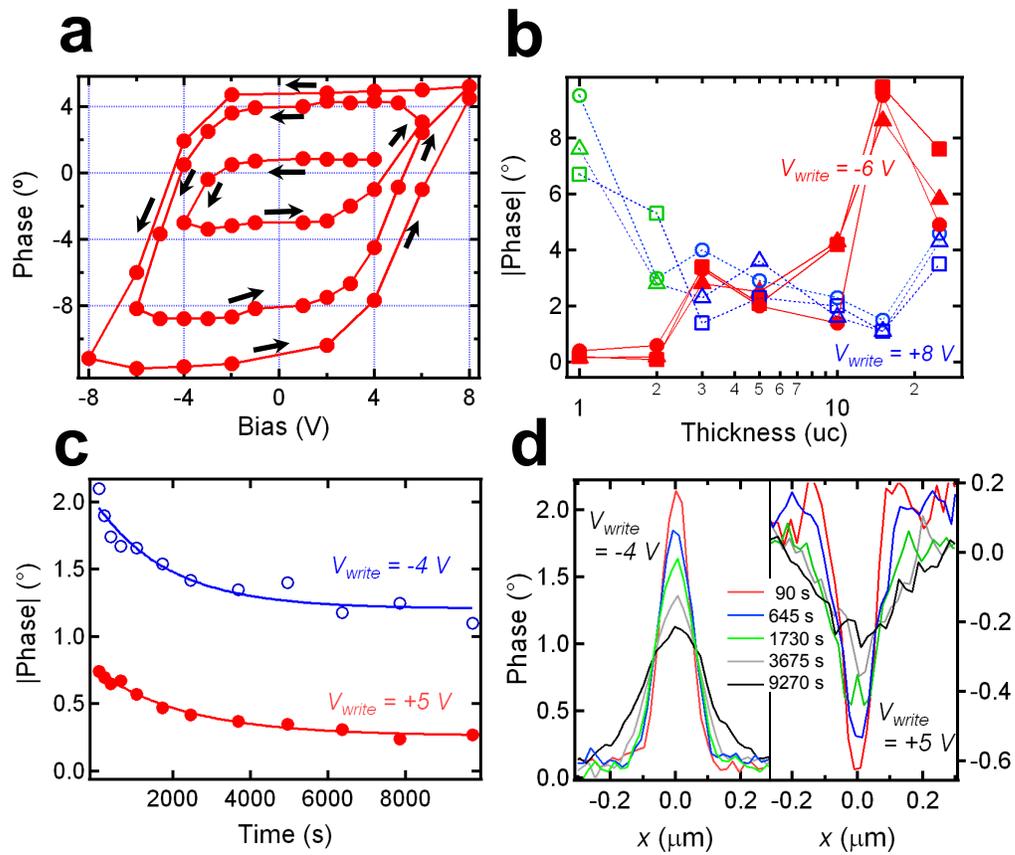

Figure 4  Y. W. Xie *et al.*

# Supporting Information

# Charge writing at the LaAlO$_3$/SrTiO$_3$ surface

Yanwu Xie,[†,‡] Christopher Bell,[†,§] Takeaki Yajima,[†] Yasuyuki Hikita,[†] and Harold Y. Hwang[†,§,*]

[†]Department of Advanced Materials Science, University of Tokyo, Kashiwa, Chiba 277-8561, Japan

[‡]State Key Laboratory of Metastable Materials Science and Technology, Yanshan University, 066004, Qinhuangdao, P. R. China

[§]Japan Science and Technology Agency, Kawaguchi, 332-0012, Japan

*To whom correspondence should be addressed, email: hyhwang@k.u-tokyo.ac.jp

## I. Materials and Methods

The LaAlO$_3$/SrTiO$_3$ samples were fabricated by epitaxially growing LaAlO$_3$ on TiO$_2$ terminated (001) insulating SrTiO$_3$ substrates by pulsed laser deposition using a KrF laser. Before growth, the substrates were preannealed at 1223 K for 30 minutes in an oxygen environment of 0.67 mPa. Following this anneal, the growth was performed at 1073 K in an oxygen pressure of 1.33 mPa, at a repetition rate of 2 Hz. The total laser energy was 20 mJ, and the laser was imaged to a rectangular spot of area approximately 2.3 × 1.3 mm$^2$ on the single crystal LaAlO$_3$ target using an afocal zoom stage. After growth, the samples were cooled to room temperature with the chamber filled by 4×10$^4$ Pa of oxygen, with one hour pause at 873 K. This procedure is the same as used elsewhere,[S1] and the oxygen annealing step is identical to Thiel et al.[S2] and Cen et al.[S3, S4] The LaAlO$_3$ film thickness were 1, 2, 3, 5, 10, 15, and 25 unit cells (uc), as measured using *in-situ* reflection high-energy electron diffraction.

A multimode atomic force microscope (AFM) Digital Instruments NANOSCOPE 3100 equipped with a NanoScope IV controller was used throughout this work. All AFM experiments were performed in air, at room temperature, with a relative humidity in the range 20 - 40%. The writing scheme is shown in Figure 1f of the main text. The conducting LaAlO$_3$/SrTiO$_3$ interface (q2DEG) is grounded by breaking through the top insulating LaAlO$_3$ layer using an ultrasonic wire bonder with Al wire. The same arrangement was also used when writing on the bare LaAlO$_3$ and SrTiO$_3$ substrates. The same PtIr5 coated silicon tip (Arrow NCPT, Nanoworld. Force constant = 40 N/m, tip radius of curvature ≤ 25 nm) is used for both writing and reading. All of the writing was performed in tapping-mode with amplitude feedback. Similar writing (not shown) could also be achieved using contact-mode.

## II. Writing and erasing conductive paths on the $d_{LAO}$ = 3 uc sample

Al electrodes were defined by *ex-situ* optical lithography and lift-off evaporation of a ~80 nm thick layer of Al, after first etching the contact pads to a depth of ~30 nm using $Ar^+$ ion milling. The gap between the Al electrodes, which are connected to an external electrometer, is ~15 μm. Each of the Al electrodes were then contacted by a series of wires written on the interface using $V_{write}$ = +8 V. In this way an electrode was formed in the LAO/STO film, free from the large height variations found close to the Al electrodes. This structure then formed the electrode for further writing (see Figure 1 in the main text). Before writing, the sample was kept in a dark environment for more than 10 hours to suppress photo-excited carriers in the bulk $SrTiO_3$.[S3,S4] A constant current of 1 nA was applied through the two electrodes and the voltage needed to maintain the current was monitored.

## III. AFM parameters

The phase variation of the written features depends sensitively to the status of the AFM tip, the biases applied for both writing and reading, the amplitude setpoint applied to the tip, and the lift height. A small value of the amplitude setpoint means a large force applied to the sample. A typical value of the amplitude setpoint for normal imaging is about 1 V. Table SI summarizes the parameters used in experiments.

Table SI. Summary of parameters used in writing and reading.

| Figure | Writing | | | Reading | |
|---|---|---|---|---|---|
| | Tip Bias, $V_{write}$ (V) | Amplitude set point (V) | Tip speed (μm/s) | Tip Bias, $V_{read}$ (V) | Lift height (nm) |
| 1 | Various (see Fig. caption) | 0.02 | 0.3 | -2 | 20 |
| 2a | -5 +5 | 0.02 | 0.3 | +2 | 20 |
| 2b | | 0.02 | 0.1 | -2 | 20 |
| 2c | | | | N/A | 20 |
| 2d | | | | -5 | 0 |
| 3a | -8 to +8, 2 V step | 0.02 | 0.3 | +2 | 20 |
| 3b | | | | -2 | 20 |
| 3c | | | | 0 | 0 |
| 4a | Various (see Fig. caption) | 0.01 | 0.2 | +5 | 20 |
| 4b | -6 | 0.02 | 0.2 | -2 | 20 |
| | +8 | 0.01 | 0.3 | -2 | 20 |
| 4c&4d | -4 | 0.03 | 0.2 | -2 | 20 |
| | +5 | 0.02 | 0.2 | | |
| S1a,S1c,S1e,&S1g | -8 | 0.01 | 0.3 | -2 | 20 |
| S1b,S1d,S1f,&S1h | +8 | 0.01 | 0.3 | -2 | 20 |

## IV. Exponential decay fitting

Table SII lists the fitting parameters obtained for the lines in Fig. 4c of the main text.

Table SII. The parameters of the fitted curves shown in Fig. 4c in the main text, of the form $p = p_0 + p_1 \exp(-t/\tau)$.

| Curve | $p_0$ (°) | $p_1$ (°) | $\tau$ (s) |
|---|---|---|---|
| -4 V | 1.20 ± 0.06 | 0.79 ± 0.07 | (2.0 ± 0.5)×10$^3$ |
| +5 V | 0.26 ± 0.02 | 0.49 ± 0.02 | (2.4 ± 0.3)×10$^3$ |

## V. Charge writing on bare (001) LAO and (001) STO substrates

Figure S1 shows EFM images of charges written on bare LAO and STO substrates. As noted in the main text, the charges are relatively more stable on LAO, as compared to STO.

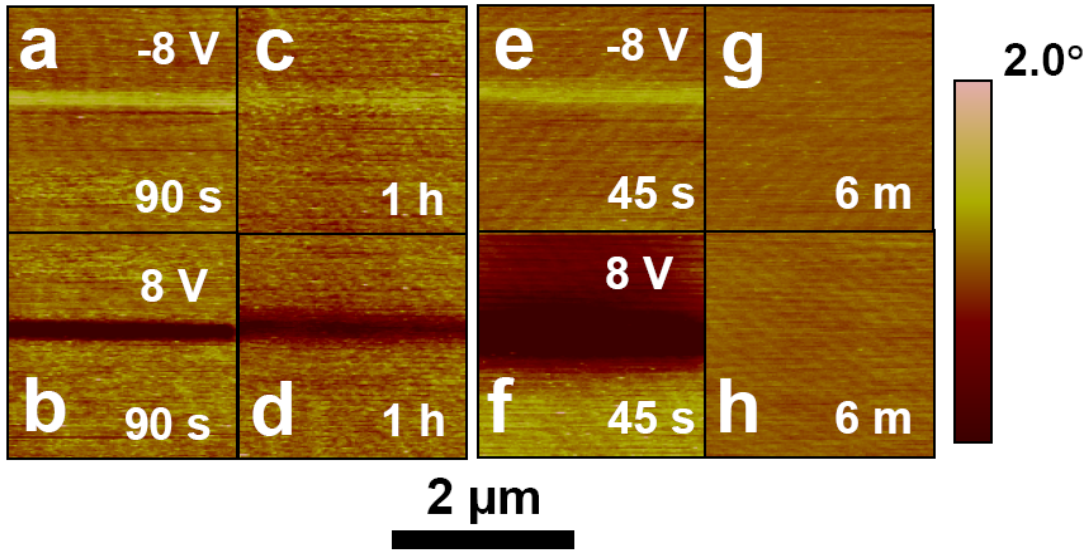

FIGURE S1. Charge writing on bare (001) LAO and (001) STO substrates. Written features on LAO with (a), -8 V (b), +8 V acquired after 90 seconds. (c, d) The same areas as (a) and (b), respectively, acquired after 1 hour. Features on STO, written with (e) -8 V (f) +8 V acquired after 45 seconds. (g, h) The same areas as in (e) and (f), respectively, acquired after 6 minutes. All images were acquired with $V_{read}$ = -2 V.

**References:**
S1. Bell, C.; Harashima, S.; Kozuka, Y.; Kim, M.; Kim, B. G.; Hikita, Y.; Hwang, H. Y. *Phys. Rev. Lett.* **2009**, 103, 226802.
S2. Thiel, S.; Hammerl, G.; Schmehl, A.; Schneider, C. W.; Mannhart, J.; *Science* **2006**, 313, 1942-1945.
S3. Cen, C.; Thiel, S.; Hammerl, G.; Schneider, C. W.; Andersen, K. E.; Hellberg, C. S.; Mannhart, J.; Levy, J. *Nature Mater.* **2008**, 7, 298-301.
S4. Cen, C.; Thiel, S.; Mannhart, J.; Levy, J. *Science* **2009**; 323, 1026-1030.